\documentclass[english,sigconf]{acmart}

\usepackage{color}
\usepackage{xspace}
\usepackage[inline]{enumitem}
\usepackage{hyperref}
\usepackage{listings}

\definecolor{red}{rgb}{1,0,0}
\definecolor{green}{rgb}{0,1,0}
\definecolor{blue}{rgb}{0,0,1}
\definecolor{cyan}{rgb}{0.4,1,1}
\definecolor{orange}{rgb}{1,0.7,0}
\definecolor{dkgreen}{rgb}{0,0.6,0}
\definecolor{gray}{rgb}{0.5,0.5,0.5}
\definecolor{purple}{rgb}{0.58,0,0.82}
\definecolor{orange}{rgb}{1,0.45,0.13}        
\definecolor{olive}{rgb}{0.17,0.59,0.20}
\definecolor{brown}{rgb}{0.69,0.31,0.31}

\AtBeginDocument{%
  }

\copyrightyear{2023}
\acmYear{2023}
\setcopyright{acmlicensed}\acmConference[WWW '23 Companion]{Companion Proceedings of the ACM Web Conference 2023}{April 30-May 4, 2023}{Austin, TX, USA}
\acmBooktitle{Companion Proceedings of the ACM Web Conference 2023 (WWW '23 Companion), April 30-May 4, 2023, Austin, TX, USA}
\acmPrice{15.00}
\acmDOI{10.1145/3543873.3585572}
\acmISBN{978-1-4503-9419-2/23/04}

\acmConference[WWW'23]{The Web Conference}{April 30--May 4, 2023}{Austin, Texas, US}

\begin{document}

\title{Caught in the Game: \\On the History and Evolution of Web Browser Gaming}

\author{Naif Mehanna}
\email{naif.mehanna@univ-lille.fr}
\affiliation{%
  \institution{Univ. Lille, CNRS, Inria}
  \city{Lille}
  \country{France}
}

\author{Walter Rudametkin}
\email{walter.rudametkin@irisa.fr}
\affiliation{%
  \institution{Univ. Rennes, CNRS, Inria, \\Institut Universitaire de France (IUF)}
  \city{Rennes}
  \country{France}
}

\renewcommand{\shortauthors}{Mehanna and Rudametkin}

\begin{abstract}  
    Web browsers have come a long way since their inception, evolving from a simple means of displaying text documents over the network to complex software stacks with advanced graphics and network capabilities.
    As personal computers grew in popularity, developers jumped at the opportunity to deploy cross-platform games with centralized management and a low barrier to entry.
    Simply going to the right address is now enough to start a game.
    From text-based to GPU-powered 3D games, browser gaming has evolved to become a strong alternative to traditional console and mobile-based gaming, targeting both casual and advanced gamers.
    Browser technology has also evolved to accommodate more demanding applications, sometimes even supplanting functions typically left to the operating system.
    Today, websites display rich, computationally intensive, hardware-accelerated graphics, allowing developers to build ever-more impressive applications and games.
    
    In this paper, we present the evolution of browser gaming and the technologies that enabled it, from the release of the first text-based games in the early 1990s to current open-world and game-engine-powered browser games.
    We discuss the societal impact of browser gaming and how it has allowed a new target audience to access digital gaming.
    Finally, we review the potential future evolution of the browser gaming industry.    
\end{abstract}



\pagestyle{plain}
\settopmatter{printfolios=true}
\maketitle

\section{Introduction}
\label{sec:intro}

The web as we know it today is the result of a steady evolution.
Accessing the internet allows users to perform a wide array of activities, including browsing interactive web pages, watching movies, or playing competitive video games directly in the browser.
However, features that are considered mainstream today are the result of over 30 years of web browser evolution.

In 1990, Tim Berners-Lee created \textit{WorldWideWeb} (later renamed to \textit{Nexus}), which is seen as the first web browser ever created. \textit{WorldWideWeb} enabled non-technical users to access \textit{hypermedia} content through \textit{hypertext}.
This first version of the web did not include any graphic content, as this was not its intended purpose.
Instead, the \texttt{WorldWideWeb} navigator was initially designed to edit and display simple text-based documents on the network.
While the documents it supported included styling, their impact on the page was limited and the structure of their styling files was significantly different from current \textit{Cascading Style Sheet}\footnote{\url{https://www.w3.org/Style/CSS/Overview.en.html}} standards.

In 1994, the first browser to natively support embedded graphics was released.
\textit{Mosaic}\cite{mosaicBrowser} allowed images to be displayed side-by-side with text, while being easier to install and launch than Nexus. This allowed Mosaic to reach a significantly larger user base.
The browser rapidly evolved to form the basis of \textit{Netscape Navigator}, which introduced the first version of a scripting language (initially named Mocha, then LiveScript, and finally JavaScript) in 1995~\cite{keith2005brief}, enabling more interactivity in web pages.
However, the initial version of JavaScript, which shipped with Netscape 2.0~\cite{netscape2.0b}, lacked many of the features intended by its creator, Brendan Eich, when he first prototyped the scripting language.
The absence of advanced features was due to the scripting language not being considered a priority for the new version of Netscape Navigator. Furthermore, the language exhibited various bugs that made it relatively unsuitable for advanced use~\cite{brock2020javascript}.
JavaScript's initial built-in library included a set of general-purpose objects that permitted interaction with the HTML document, setting the base for the first version of the \textit{Document Object Model, level 0} (DOM). JavaScript 1.0 gained traction with web developers, and JavaScript 1.1 quickly followed in the next major release of Netscape Navigator (3.0) with improved performance and features.

In 1995, the release of Internet Explorer 1.0 by Microsoft and the subsequent release of Internet Explorer 2.0 three months later as freeware (unlike its competitors) started what is now known as "the first browser war"~\cite{chiaravutthi2006firms}. As both browsers competed for market share, new features were rapidly introduced.
Netscape's release of JavaScript prompted Microsoft to announce the release of its own implementation of the scripting language in Internet Explorer 3.0: JScript. JavaScript and JScript later became implementations of the ECMAScript standard, although they supported different sets of features due to the frequent feature releases of both browsers.
Although JavaScript was maturing, developing dynamic content was still challenging due to compatibility issues between the two leading browsers. Content that was optimized for one browser could look different on the other, leading web developers to prioritize one browser at the expense of the other.
By 1996, browsers were evolving rapidly, with customization options becoming available. This was exemplified by Internet Explorer 3.0, which introduced support for \textit{Cascading Style Sheets} (CSS), with Netscape following suit in its fourth iteration.
With the \textit{World Wide Web Consortium (W3C)} recommending CSS, HTML being established as the standard document format, and the definition of the ECMAScript standard, web developers gained the tools needed to transition from static web pages to dynamic and reactive websites.
These advancements paved the way for the emergence of browser-based games. Starting as a limited platform for game development, web browsers evolved to include complex game mechanics, such as multiplayer games that relied on real-time interactions between players, becoming ideal platforms for reaching a growing number of players. Browsers made it easy for both amateur and professional game developers to distribute their games.

In this paper, we present the history and evolution of browser-based games. In Section~\ref{sec:prehistory}, we explore how the lack of features in early browsers only allowed games that did not require advanced interactivity.
Section~\ref{sec:middleage} presents the fast-paced evolution of the web and support for third-party plug-ins in browsers, such as Flash and Java, which allowed a significant leap forward for browser game development.
Section~\ref{sec:nowadays} dives into the capabilities offered by HTML5 and its native support of 2D and 3D graphics, enabling the development of cross-platform rich games directly in the browser.
Finally, in Section~\ref{sec:discussion}, we discuss the impact and future of browser games.

\section{Early Browser Gaming}
\label{sec:prehistory}

The emergence of consumer-friendly web browsers in the mid-1990s allowed for the rise of game-related websites. However, early browsers had limited features and standards, and their lack of client-side scripting made interactive experiences slow, especially for users with dial-up modems.
These limitations made the development of browser-based games challenging.

Text-based games, which had been around since the early days of terminal-based computers, were able to take advantage of the limited features of early browsers. Despite their simple interfaces, text-based games attracted a significant number of players and initiated the era of browser games.

As web browsers evolved rapidly, web developers worked on standardizing and making new features accessible. Client-side scripting, in the form of JavaScript and the ECMAScript standard, was introduced by Netscape 2.0 in 1995, as explained in Section~\ref{sec:intro}. Over the following years, Internet Explorer introduced its implementation of client-side scripting (JScript) and the first standardized version of Cascading Style Sheets (CSS).

These new tools allowed the emergence of new types of browser games that took advantage of increased interactivity allowed by JavaScript, its Document Object Model (DOM) APIs, and standardized styling. In this section, we explore how text-based games took advantage of the early browsers' limited capabilities and how newly introduced features of web browsers enabled the emergence of dynamic games using Dynamic HTML (DHTML).

\subsection{Interactive fiction}

Text-based games, also known as interactive fictions, have been around since the late 1960s with the advent of mainframe computers.
These games allowed players to take part in adventures by reading textual descriptions of events and entering a textual representation of their chosen actions.
The genre's popularity rose with the introduction of affordable home computers and terminal monitors that were only capable of displaying text-based content.

\begin{figure}[t]
    \includegraphics[width=0.9\columnwidth]{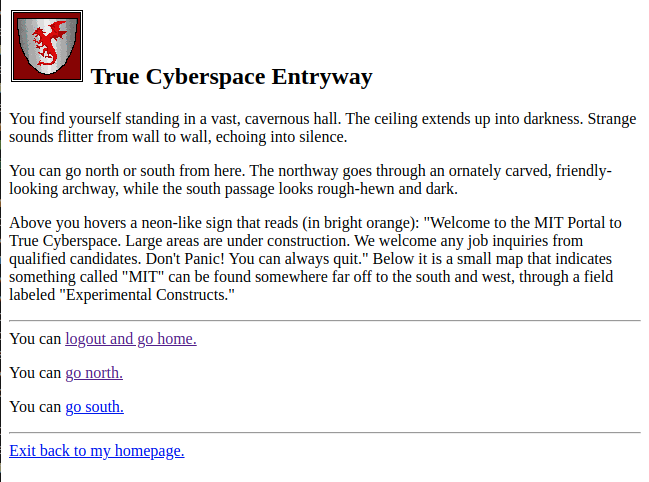}
    \caption{CyberMUD, an interactive fiction driven by hyperlinks
    %~\urlfootnote{https://bkgm.com/motif.go.html)
    }
    \label{fig:cybermud}
\end{figure}

When web browsers were first introduced, operating systems already supported advanced graphics capabilities. However, early browsers lacked the ability to display complex graphics, which made browser-based text games seem like a step back for the gaming ecosystem. This was because home computers already supported more advanced games that were comparable to console games in terms of performance. As a result, browser-based interactive fictions were less appealing to many gamers.
However, web browsers offered  online connectivity and platform-independent compatibility.
As a result, browser-based interactive fictions started to appear.
They were easily accessible through their \textit{Uniform Resource Locator (URL)}, eliminating the need to purchase and install the game.
This gave browser-based interactive fictions an edge over other gaming systems, attracting large communities of players who could share their experiences in online forums, bulletin boards, and later in social media.
Figure~\ref{fig:cybermud} displays a screen capture of \textit{CyberMUD}, an interactive fiction developed in 1994 using HTML. Players interact with the game through \textit{hyperlinks}.

Another key factor that contributed to the rise of online interactive fictions is the emergence of \textit{server-side programming}.
With the release of PHP and MySQL in 1995~\cite{stobart2004introduction}, dynamic websites became possible, allowing for more advanced user interactions. Without server-side programming, websites would have remained static and unable to react to user input, which was essential for the interactive fiction genre, and particularly for online interactive fictions.

One of the earliest successful online text-based strategy games was \textit{Earth: 2025}, created in 1996 by Mehul Patel. This game allowed players on the internet to belong to virtual countries that they managed and defended against other players, and it gathered a significant user base during its lifespan, remaining active until 2009. In 1997, \textit{Hattrick} and \textit{Alien Adoption Agency} were both released and also attracted numerous players.

While online text-based games have lost popularity due to the improved abilities of browsers, the game genre still persists to this day on various websites. \textit{TextAdventures}\footnote{\url{http://textadventures.co.uk/}} is one such website that allows current internet users to play interactive fiction.

\subsection{The DHTML era}

The introduction of client-side scripting changed the shape of the web. At the end of the 1990s, the main way to access the internet was through dial-up modems, which came with several limitations, such as low bandwidth and relatively high latency due to their 56 kbits/s speed limitation~\cite{patrick2019brief}.
Before client-side scripting was available, validating any information on the web page took a full round trip to the server, which significantly limited the user experience. Client-side scripting allowed user interactions to happen directly in the browser.
The first release of JavaScript on Netscape 2.0 was limited, only allowing text elements to be manipulated by the language, but web developers quickly saw the potential of JavaScript for web development.
Although JavaScript was initially limited to text elements, it allowed the creation of a number of games that relied on \textit{ASCII} animations running in a \textit{textarea} element. An example of such a game can be found in the black and white ASCII implementation of \textit{Space Invaders}, which ran directly in the Netscape browser and is still playable today~\cite{spaceinvadersAscii}.

The release of Netscape 3.0 and JavaScript 1.1 brought several much-needed improvements to the scripting language. One such improvement was the ability to change the source attribute of image elements, which had a significant impact on the development of browser games. Real-time animations allowed developers to create games that provided an arcade-like experience. As early as 1996, DHTML versions of the notorious \textit{Pac-man} were developed, followed by reproductions of other arcade games like \textit{Space Invaders} (non-ASCII reproductions) or \textit{SmoothMaze}.
However, most full DHTML-based games were developed for demonstration purposes and didn't achieve wide-scale distribution.

There are several reasons why DHTML games never truly took off. First, slow speeds of dial-up modems resulted in long loading times, as resources needed to be loaded before any JavaScript-based game could start. As more complex games required a higher number of assets, slow internet speeds were seen as a significant limitation to DHTML games. Second, competition with plug-in-powered games highlighted many limitations of JavaScript games. They could not compete with the animations and performance offered by Java or Flash-based games (see \S\ref{sec:middleage}). Finally, although the competition between Netscape and Internet Explorer led to many features being developed quickly, standardization efforts of the scripting language, as recommended by the W3C, were less of a priority. This made JavaScript-based game development complicated, since features might behave differently between browsers. The quick pace of browser development also introduced various breaking changes to the different APIs, making JavaScript-based games prone to breakage and incompatibility

\section{The rise in popularity of browser gaming}
\label{sec:middleage}

The release of Netscape 2.0 in 1996 introduced features~\cite{netscape2.0}, such as JavaScript, that focused on improving interactivity on the web, as discussed in Section~\ref{sec:prehistory}.
However, more notably for gaming was the release of the \textit{Netscape Plug-in API (NPAPI)} and the resulting support for Java Applets, and later, Flash applications.
The \textit{Plug-in API} allowed the browser to interact with a wide array of new content types that would have otherwise remained unavailable to web browsers.
The browser delegated content it didn't understand, such as documents, programs or multimedia, to external plug-ins if they were available.
The result was a significant improvement in the extensibility of the Netscape browser, and the availability of plug-ins such as the \textit{Java Runtime Environment}, \textit{Adobe Acrobat PDF reader}, \textit{Apple Quicktime}, and \textit{Macromedia Director} that opened up new possibilities for browsers with performance comparable to native applications.
Interestingly, the \textit{NPAPI} interface was not limited to Netscape's proprietary plugins. It also allowed third-party developers to provide their own frameworks to enhance the browsing experience.

In the next subsections, we explore the evolution of Java applets and the significant advancements they introduced to browser gaming before exploring the enduring impact of the Flash plug-in.

\subsection{Java-applet games}
\label{subsec:java}

The Java programming language was created by Sun Microsystems in 1995, only one year prior to its introduction as a plug-in in Netscape Navigator 2.0. One of Java's main advantages is that it's cross-platform; code written in Java is executed by the \textit{Java Virtual Machine (JVM)} rather than directly by the operating system. Sun Microsystems coined the slogan \textit{"Write once, run everywhere"} to market this idea. The portability of the language made it an ideal candidate for browser integration as it could be run on any operating system given where the JVM had been ported to.
Java was first introduced to the web in the \textit{HotJava Browser}~\cite{hotjavaAlpha}, developed by Sun Microsystems to promote their vision of a web built around Java applets. Netscape's support for Java applets came at a time when Java was gaining popularity and being promoted by major players in the IT industry, such as Sun Microsystems, Apple, and Microsoft.

Java applets revolutionized web development by enabling web pages to be developed as interactive applications, rather than just static pages with information.
The HotJava Browser's implementation of the JVM allowed Java classes to directly interact with the page's HTML content, unlocking the full capabilities of the host machine to modify the page's content.
Despite serious drawbacks, such as high memory usage and increased processor load, as well as a long history of security issues, the benefits that Java brought to web development arguably outweighed these limitations.

Java applets gained significant popularity with the launch of the Java NPAPI plug-in in 1996. Not long after, Microsoft joined the trend with the introduction of NPAPI plug-in support in Internet Explorer 3.0. By 1998, both major browsers supported the execution of Java applets~\cite{internetExplorer3.0}.

Java Applets started a new era of games in the browser, as interactivity is a key component of any game. Their introduction marked a significant breakthrough as they offered animations, audio, and real-time interactivity, surpassing the limitations of JavaScript. While JavaScript was too slow to render game-grade animations and its adoption remained limited in the gaming ecosystem, Java Applets paved the way for immersive browser-based gaming experiences.
Moreover, Java applets greatly reduced the burden of incompatibility between browsers, as the JVM executed Java code much more consistently across platforms.

\begin{figure}[h]
    \includegraphics[width=0.8\columnwidth]{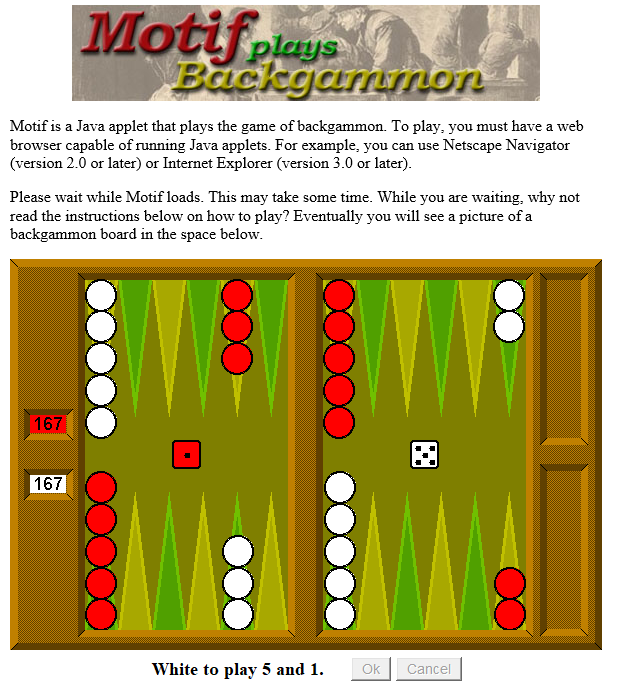}
    \caption{The Backgammon game as a Java applet
    %~\urlfootnote{https://bkgm.com/motif.go.html)
    }
    \label{fig:bckgm}
\end{figure}

\begin{figure}[t]
    \includegraphics[width=0.85\columnwidth]{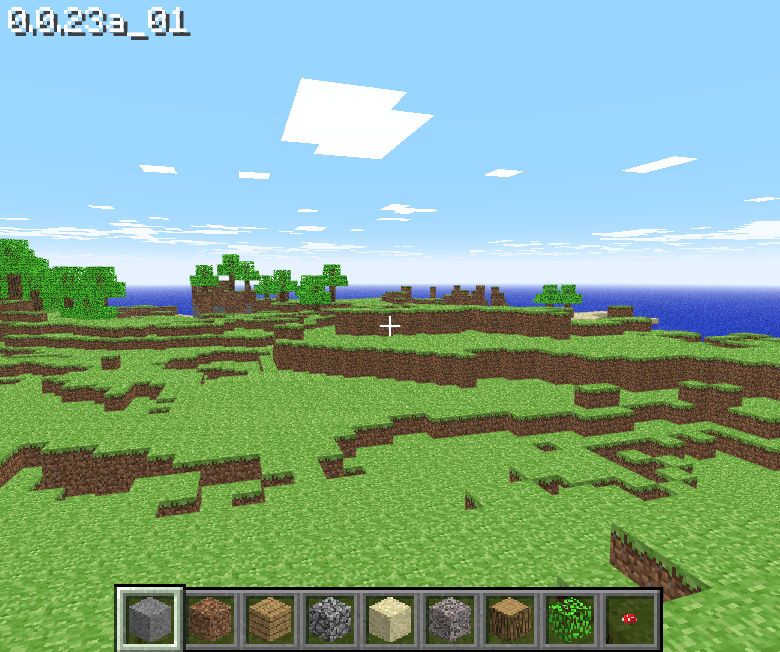}
    \caption{Minecraft Classic in the browser
    %~\urlfootnote{https://bkgm.com/motif.go.html)
    }
    \label{fig:minecraft}
\end{figure}

Java graphics were based on bitmap images, and games developed using Java Applets initially had simple graphics.
Figure~\ref{fig:bckgm} shows an example of a \textit{Backgammon} game that was available in 1998.
At that time, limited internet speeds meant games with higher resolution graphics, being larger in size, took significantly longer to load.
However, with the advent of broadband internet, which improved speeds, Java-based browser games naturally evolved to include better graphics and 3D assets. In addition, lower network latency allowed for improved online connectivity in web pages, leading to the release of several \textit{Massively Multiplayer Online Role-Playing Games (MMORPGs)}.
One of the most popular MMORPGs at the time was RuneScape,\footnote{\url{https://play.runescape.com/}} which was written in Java and ran as an applet in the browser.
The game initially included a mixture of 2D and 3D graphics before turning into a fully 3D game in later releases.
Another example of a game utilizing the capabilities of Java Applets is \textit{Minecraft}, which was released in 2009 as a 3D game with online connectivity and quickly gathered a significant user base. The game allows players to build their own world with almost no limits. Figure~\ref{fig:minecraft} shows an example of Minecraft.

While Java provided web developers with improved tools to build powerful games running in the browser, applets started to lose popularity in the early 2010s.
The official demise of applets was initiated in 2013 when Google Chrome, which had already become the most popular browser, began dropping support for the \textit{NPAPI} that most plug-ins, including Java, relied on.
There were several reasons for this, including a dependency on the JRE's version, numerous security issues, and the increased capabilities of JavaScript and Google Chrome's V8 engine, which allowed JavaScript to be compiled to machine code, thus improving its performance. Java applets were officially deprecated by Oracle in 2017 with the release of the Java Development Kit 9 (JDK9)~\cite{jdk9Deprecated}.

\subsection{Flash-based games}
\label{subsec:flash}

In 1996, \textit{FutureWave Software} created the \textit{Adobe Flash Player} (initially named \textit{FutureSplash}) to allow developers to design interactive animations through a simple and accessible interface.
\textit{Flash} was first introduced in Netscape 2.0 as part of a featured list of plug-ins following the release of its \textit{NPAPI} interface. In 1997, the plug-in was acquired by Macromedia~\cite{acquisitionFlashMacromedia}, which continued its development and extensively marketed it, leading to its quick rise in popularity. Macromedia was later bought by Adobe in 2005~\cite{adobeMacromedia}, and the plug-in was rebranded as the \textit{Adobe Flash} plug-in.

While providing cross-environment support for its graphics and animations, Flash also came with a series of advantages over Java applets that led to its dominance in the browser-gaming ecosystem. 
One of the main advantages of Flash was its utilization of vector graphics, which resulted in a reduction of storage and distribution requirements for graphics when compared to bitmap images. In the era of high latency and low bandwidth, the loading times of animations were greatly reduced~\cite{salter2014flash} in Flash applications, improving the user experience.
Another advantage of Flash was its widely supported export format (SWF), which allowed for easy sharing of the created resources.
Alongside the Flash plug-in, various tools, such as \textit{Adobe Animate},\footnote{\url{https://www.adobe.com/products/animate.html}} were released to build and develop interactive content without programming knowledge, allowing content creation to be more accessible.
By 2002, Flash was installed in over 98\% of browsers~\cite{guldman2002building} and widely used across the web.

Interestingly, initially, Flash was mainly used for building interactive and animated content for web pages. Flash did not allow users to script actions: while it provided the ability to add buttons, the components themselves could not do significantly more than skip animation frames, limiting its use for game development. Macromedia vowed to surpass this limitation by introducing \textit{ActionScript} in 2000. ActionScript was based on the ECMAScript standard and was therefore syntactically close to JavaScript, allowing web developers familiar with the latter to quickly start developing advanced interactive content for Flash.
With the introduction of ActionScript, Flash games started appearing on various websites. The first platform to allow users to upload their Flash games was Newgrounds~\cite{newsground}.
At its peak, Newgrounds had over $80,000$ games~\cite{fiadotau2020growing}, virtually every game genre, ranging from adventure games to educational games.
In a time when social media platforms were just beginning, Newgrounds provided users with a platform to express themselves. As such, various games were built as a reflection of current political events, such as the famous \textit{Bad Dudes vs. Bin Laden} game~\cite{fiadotau2020growing}, developed in reaction to the \textit{9/11} attacks.

\begin{figure}[t]
    \includegraphics[width=0.8\columnwidth]{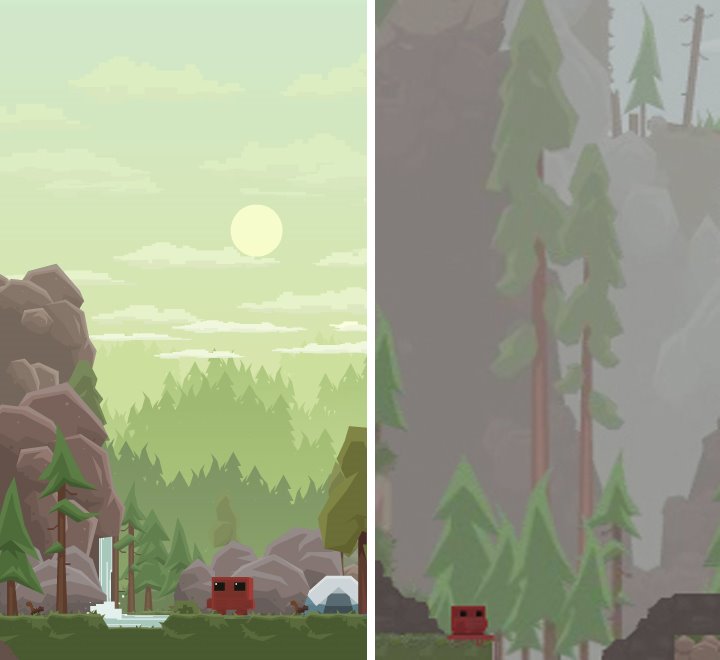}
    \caption{A comparison between the Wii (left) and Flash (right) versions of Super Meat Boy~\cite{supermeatboy}.
    %~\urlfootnote{https://bkgm.com/motif.go.html)
    }
    \label{fig:supermeat}
\end{figure}

Flash game development became more accessible, leading to an explosion of the number of games created. Although it is impossible to quantify the exact number of Flash games developed during its 24-year lifespan, Flashpoint~\cite{flashpoint}, a game preservation project, currently lists over $100,000$ Flash games in its database, making it the largest game library ever created for a single platform. For comparison, the \textit{Playstation 2} had $3,874$ games developed for it during its lifespan.
The accessibility of Flash game development also gave rise to the indie game scene, with games originally created in Flash later being ported to various consoles. For instance, \textit{Super Meat Boy} (shown in Figure~\ref{fig:supermeat}), whose Flash graphics style is still recognizable in current versions of the game, and \textit{Canabalt}, which is playable on various consoles such as the \textit{Playstation 3} and the \textit{Playstation Portable}.

Flash's popularity peaked during the first decade of the 2000s, with many high-impact games created during this time. \textit{Stardoll}~\cite{stardoll}, a fashion game that focused on its online community, had over 400 million users as of 2016, while \textit{Neopets}~\cite{neopets} (1999), where users care for virtual animals, persists to this day in Facebook's metaverse. \textit{FarmVille}~\cite{farmville}, launched in 2009 on Facebook's game platform, attracted over 34 million daily users who managed their virtual farms and quickly became one of the highest-grossing games on Facebook's platform.

However, Flash's advantages were eventually outweighed by its disadvantages, including security issues introduced by the plug-in~\cite{buhov2018flash}. Additionally, the growing adoption of smartphones, especially Apple's iPhone, which did not support Flash due to closed-source code, high CPU usage, and extensive security issues as cited in their 2010 blog post \textit{"Thoughts On Flash"}~\cite{thoughtsflash}, led to its decade long decline. In July 2017, with the Flash install-base representing only a fraction of its peak, Adobe announced the scheduled end of Flash by 2020~\cite{endoflife}.
New web standards have since been introduced, covering much of the technical landscape that Flash was conceived for with, arguably, less drawbacks including being directly provided by the browser.

\section{HTML5 and the advent of the canvas}
\label{sec:nowadays}

Over the years, the HTML markup language has evolved multiple times in an attempt to provide more advanced standardized features. The first major change to HTML was in 1997, with the release of HTML 4.0~\cite{w3chtml4}, which introduced official support for CSS documents and dynamic scripting using JavaScript. Web usage continued to grow, with many users adopting the internet and increasingly relying on it to access information, purchase goods, or for entertainment. The web required more interactivity, and Adobe Flash and Java applets were created as solutions to provide dynamic content (\S\ref{sec:middleage}). However, they relied on external plug-ins and lacked standardization.
Despite its limitations, the web was quickly becoming more dynamic, and many web pages began providing interactive content. Many websites could be compared to native applications in their complexity and features. In response, an early draft for \textit{Web Applications 1.0} was published by the \textit{Web Hypertext Application Technology Working Group (WHATWG)} in 2005~\cite{webApps2005}. After years of conjoint work between the \textit{WHATWG} and the \textit{W3C}, the Web Application 1.0 specification evolved in 2008 into what is known today as HTML 5.0, an official standard of the W3C. This was an important turning point, as web applications began to achieve features comparable to native applications.
HTML5 provided many advantages. Native video and audio support allowed websites to distribute multimedia content without relying on external plug-ins, all while supporting more formats~\cite{hawkes2011foundation}. The \textit{canvas} element was officially introduced, allowing web pages to natively integrate 2D graphics and permitting game engines based on HTML5 to emerge. Both of these new additions were enough to make the Adobe Flash plug-in obsolete.

In the following subsections, we cover the capabilities enabled by the \textit{canvas} element for browser gaming. We will then cover the official support of \textit{WebGL}, which allows hardware-accelerated 3D graphics to be incorporated directly into the browser, paving the way for an entirely new dimension of browser games. Finally, we will explore cloud-based browser gaming, made possible by improved video capabilities and the adoption of fiber optic broadband, which significantly increases bandwidth and reduces latency.

\subsection{The Canvas element}
\label{subsec:canvas}

Although the canvas element became a standard element of HTML5, it actually originated as a non-standard element in the Webkit engine,\footnote{\url{https://webkit.org/}} which was part of the Safari browser back in 2004~\cite{canvasWebkit2004}, four years before the official release of the HTML 5.0 specifications.
Before the introduction of canvas, most browsers natively supported bitmap images, \textit{Scalable Vector Graphics (SVG)} images, and gradients.
Notably, Internet Explorer did not support SVG and instead supported the \textit{Vector Markup Language (VML)} which never really took off.

The canvas element allowed browsers to draw graphics in real-time and could be controlled directly through JavaScript, which also has access to the page's DOM. This provided greater interactivity and animations without the need for external plug-ins. The canvas element was supported by all HTML5 compliant browsers, including mobile devices. However, as rendering frames was dependent on the underlying JavaScript code, canvas-based animations relied on JavaScript's performance. In its early stages and up until the introduction of the V8 engine by Google Chrome, JavaScript was not known for its execution speed.
The V8 engine significantly improved JavaScript performance and helped establish the scripting language as a solution for fast web applications.
In 2012, hardware-acceleration was added to 2D canvas-based graphics~\cite{hardwareAcc}.
This significantly improved drawing times and helped avoid CPU bottlenecks.
The drawing instructions can be offloaded to the GPU, if available and supported, drastically increasing rendering speeds.
These advances made canvas-based graphics more attractive to game developers, and canvas-based game engines started to appear.
These engines provided tools that developers needed to build their games, such as physics engines, making browser game development faster and simpler.
This bridged some of the gap between browser-based games and native games while remaining cross-platform, centralized and not requiring users to install the games.

\begin{figure}[t]
    \includegraphics[width=0.9\columnwidth]{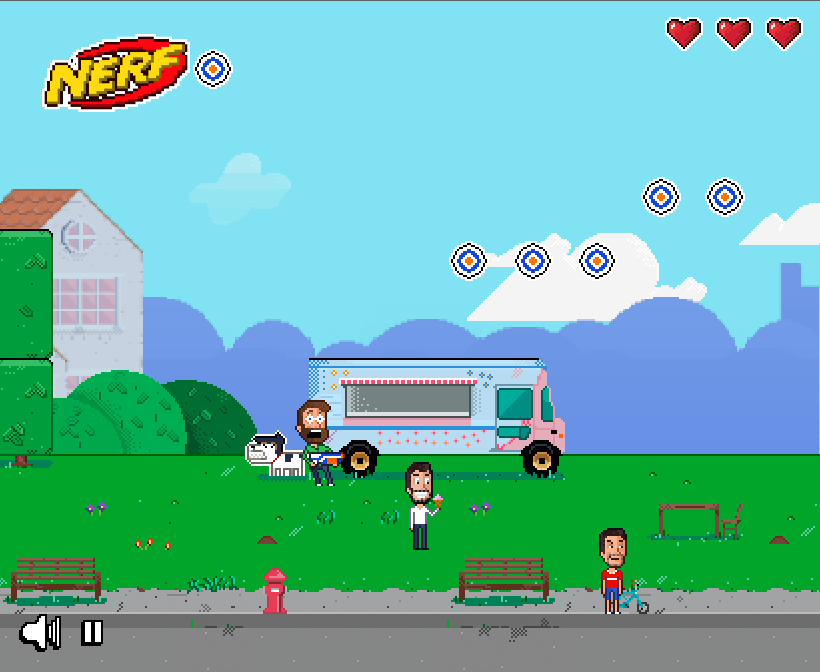}
    \caption{Be A Hero Again, a 2D canvas game developed with Phaser.
    %~\urlfootnote{https://bkgm.com/motif.go.html)
    }
    \label{fig:nerfphaser}
\end{figure}

\textit{Phaser}~\cite{phaser} is an example of a popular 2D game framework that was developed for HTML5.
The framework uses JavaScript, making it accessible to web developers that are already familiar with the language.
To date, Phaser powers a significant number of browser games that range in popularity.
Figure\ref{fig:nerfphaser} shows a screenshot of \textit{Be A Hero Again},\footnote{\url{https://beaheroagain.com}} a game developed with the Phaser framework.
The game features cartoonish, colorful graphics that are rendered using the canvas element.
\textit{MelonJS}\footnote{\url{https://melonjs.org/}} is an open-source game engine that provides the tools to build a 2D canvas-based game, such as a physics engine and support for various inputs and sensors.

It's worth noting that the improvements in graphics and speed due to the HTML5 standards and the arrival of 2D game engines for the canvas element brought browsers closer in line with what was provided by the 2D graphics of the Adobe Flash plug-in years before.

\subsection{Hardware acceleration in the browser}
\label{subsec:webgl}

Although native canvas capabilities allowed users to create interactive 2D games, PC and console-based games both offered 3D graphics. Furthermore, even recent versions of the Adobe Flash plug-in allowed developers to create GPU-powered 3D games. There was a need to introduce a standard for 3D rendering directly into the browser.
In 2006, Vladimir Vukićević laid the groundwork for the first prototype of the \textit{OpenGL 3D context} in the canvas element and later turned to the Khronos group~\cite{khronos} to create \textit{WebGL}. WebGL 1.0 was officially released in 2011 and was supported by all major browsers. It added support for native 3D graphics in the browser and used the OpenGL ES 2.0 rendering API, a subset of the native OpenGL API for embedded systems. Games created with WebGL could provide a level of graphics performance that was on par with desktop games. The significant advantage of WebGL lies in the fact that it is a standardized feature of HTML5, making it possible to exploit 3D capabilities on every platform supported by compatible browsers. WebGL 2.0 was officially introduced a few years later, in 2017, supporting new APIs and based on the more recent OpenGL ES 3.0.

WebGL is an interface to interact with the OpenGL rendering API.
For game development, WebGL represents a significant advancement, as it allows for the creation of 3D games with graphics that are on-par with desktop games. This simplifies the process of obtaining and installing a game in order to play it and attracted the interest of developers.
However, writing web applications directly in WebGL is complicated due to its low-level nature, leading to the emergence of libraries and game engines that provide higher-level abstractions.
Many major game engines now support WebGL. For example, Unity~\cite{unityGE} and the Unreal engine~\cite{unrealGE} both offer support for 2D and 3D browser games based on WebGL. Notably, Unity previously provided support for 3D browser gaming through its Unity Web Player plug-in, but lost this capability when NPAPI ceased to be supported~\cite{endNpapiUnity}.

\begin{figure}[t]
    \includegraphics[width=0.87\columnwidth]{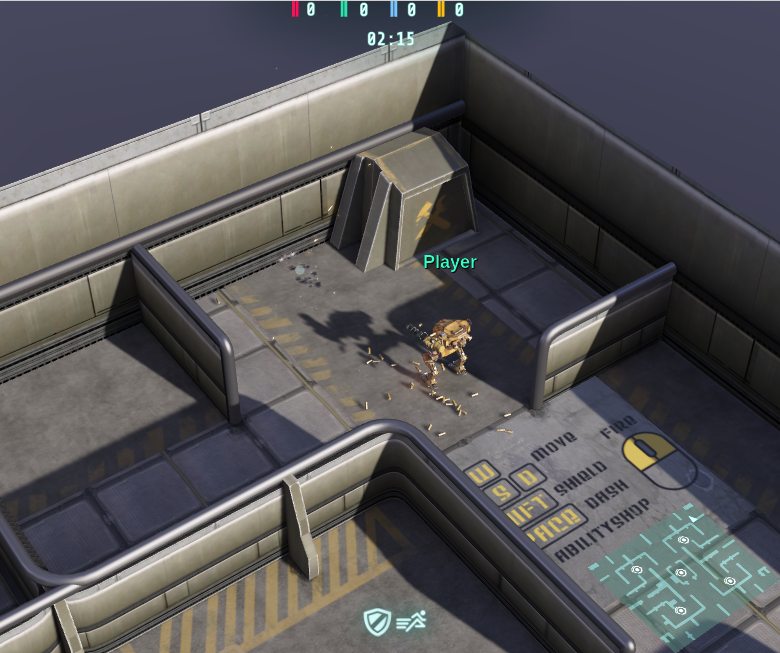}
    \caption{RoboStorm, a WebGL powered browser game.
    }
    \label{fig:robostorm}
\end{figure}

In addition to the established game engines that were ported to WebGL, others were built specifically for the development of WebGL-based games.
PlayCanvas,\footnote{\url{https://playcanvas.com}} for instance, is a popular example and provides a fully capable editor (PlayCanvas Editor) for developing in-game 3D models and interactions. Being fully written in JavaScript, PlayCanvas has the advantage of not requiring compilation, as is often the case with other established game engines that primarily use the C++ language for development.
Robostorm,\footnote{\url{https://robostorm.io}} as shown in Figure~\ref{fig:robostorm}, is an example of a 3D game that was developed with PlayCanvas. It offers online connectivity and advanced 3D graphics directly in the browser.

The popularity of WebGL is growing and new standards for browser-based 3D graphics are being developed. The \textit{WebGPU}~\cite{webgpu} specification, set to replace the WebGL standard, is currently in the working draft phase by the W3C. \textit{WebGPU} is covered in greater detail in Section~\ref{subsec:future}.

\subsection{Cloud-based gaming}
\label{subsec:cloud}

Cloud gaming has emerged as a competitive way to distribute games. Initially presented as a concept by \textit{G-Cluster} at the 2000 E3, cloud-based gaming gained in popularity towards the late 2010s as high-speed internet adoption grew among gamers, making input latency acceptable~\cite{chen2011measuring}.
Multiple platforms have proposed cloud-based gaming~\cite{cai2016survey}, such as Onlive and Gaikai, which was ultimately acquired by \textit{Sony}~\cite{sony} to form \textit{PlayStation Now} and \textit{PlayStation Plus} subscription services.
Cloud gaming has made many popular AAA games available in the browser. 
However, not all cloud gaming companies have provided browser-based access to their services~\cite{cai2016survey}.

Cloud gaming overcomes issues developers face when porting to WebGL.
For example, porting native OpenGL games to WebGL can be problematic because WebGL relies on the OpenGL ES subset API for embedded systems.
Porting from other APIs requires extensive rewrites.
Instead, cloud-based games do not run directly in the browser, they are streamed from powerful servers thanks to the video and audio elements in HTML5, side-stepping portability and hardware performance issues.
Players only need high-speed connectivity and enough processing power to decode streamed content.
Cloud gaming in browsers is made possible by the support of video and audio elements in HTML5.
Nevertheless, despite the advantages, cloud gaming suffers from a relatively limited adoption, arguably due to:

\begin{enumerate}
    \item High-speed internet access remains highly disparate. Some regions have fiber internet access, while many others still rely on ADSL~\cite{internetaccess}. Cloud gaming requires low-jitter, low-latency, high-speed networks to be effective. Limited internet access reduces the reach of the system.
    \item The upcoming WebGPU specification promises better performance, and with the growing processing power of portable devices, WebGPU might overcome the limitations posed by WebGL. A better 3D API in the browser, with more powerful devices, could reduce the appeal of cloud gaming.
\end{enumerate}

Currently, two of the most popular browser-compatible cloud-gaming services, \textit{Nvidia's GeForce NOW} and \textit{Microsoft's Xbox Game Pass}, announced a total of 20 million players each in 2022~\cite{earningsnvidia, earningsmicrosoft}.
\section{Discussion}
\label{sec:discussion}

\subsection{Impact of browser gaming}
\label{subsec:societal}

The widespread adoption of browser-based gaming early in the internet's history can be attributed to various reasons. Games for browsers were optimized for the low internet speeds of the era, resulting in an overall positive gaming experience.
Flash's vector graphics, for instance, allowed games to load quickly, even on dial-up modems, making them accessible to a high number of players.
As the internet grew and more people purchased computers for general use, browser games were typically able to run on them using the limited amounts of processing power available, making them widely accessible.
Additionally, most browser games were offered free of charge on platforms such as \textit{Newgrounds} or \textit{Armor Games}, which provided vast catalogs of free-to-play games.
The price of high-end gaming computers and home consoles often made them inaccessible, whereas browser gaming appealed to a wide audience.
It also required little setup beyond installing a plug-in, which was often pre-installed with the browser.
Browsers proved to be ideal platforms for casual gaming.

The rise of video games on the browser was not limited to single-player casual gaming. Many \textit{MMORPG}s emerged, connecting players from different regions of the world. \textit{RuneScape}, mentioned in Section~\ref{sec:middleage}, was a highly popular browser-based MMORPG that gathered millions of users, some of whom would play for hours on end. Another online MMORPG that attracted over 10 million monthly players was \textit{Habbo Hotel},\footnote{\url{https://habbo.com}} where users played in a virtual hotel.

In Section~\ref{subsec:flash}, we discussed how Flash games were a significant precursor to the indie game scene, allowing small game creators to develop popular games that were monetized and ported to multiple platforms. \textit{Super Meat Boy}, for instance, sold millions of copies when it first debuted on the \textit{Xbox} platform. Moreover, \textit{Minecraft}, which originated as a browser game (\S\ref{subsec:flash}), ranks among the most popular games of all time according to a 2021 ranking by \textit{HP}~\cite{bestSellingHP}.
Flash was accessible and straightforward, allowing even novice developers to create games and animations with relative ease. This simplicity enabled users to express themselves through Flash creations, especially games, many of which satirized political situations when social media platforms were not yet as widespread as they are today.
 
Web games span a wide range of genres, including educational, and even, adult-oriented games. Educational games, in particular, have the potential to foster new learning strategies by creating interactive and educational content that can be distributed over the internet and reach a broad audience. Browser games can also benefit scientific research~\cite{curtis2014online}, as seen in examples such as \textit{Stall Catchers},\footnote{\url{https://stallcatchers.com/}} aimed at benefiting medical research, and Phylo,\footnote{\url{https://phylo.cs.mcgill.ca/}} designed to decipher human DNA.

Today, the gaming industry is bigger than both the movie and music industries combined~\cite{gameindustry}, showing that video games are an integral part of our entertainment culture. A significant part of this success can be attributed to the rise of social games on Facebook, which gained popularity due to their simple mechanisms that attracted many first-time gamers, as well as their use of the social graph and asynchronous play~\cite{whitson2011social}. These features allowed users to play at their own pace while still sharing their gaming experience with friends, further contributing to the popularity of social games and browser gaming in general.

\subsection{The future of browser-based games}
\label{subsec:future}

Browser-based games have come a long way since the release of the early text-based games, and they continue to evolve. In 2019, the \textit{Unreal Engine 4} dropped native support for HTML5 and instead provided an extension to their engine~\cite{unrealHtml5} to cross-compile the C++ code to JavaScript.
This move signaled a disinterest in WebGL-based games.
As WebGL is based on the OpenGL API, its limitations are starting to show as the world moves towards more general-purpose uses of GPUs, such as crypto-mining or training neural networks. Thus, there is currently a need to develop a new up-to-date standard that better exploits modern GPU APIs.
One such standard is \textit{Vulkan}~\cite{vulkan}.

The upcoming \textit{WebGPU} standard is based on three current GPU APIs being developped: \textit{Vulkan} by the Khronos Group, \textit{Metal} by Apple~\cite{metal}, and \textit{Direct3D 12} by Microsoft~\cite{directx}, which provide significant advantages over OpenGL, on which WebGL is based.
In their work, Lujan~\textit{et~al.\xspace}~\cite{lujan2019evaluating} show that the Vulkan API significantly outperforms OpenGL in various scenarios.
Therefore, WebGPU is expected to open a new era for 3D games in the browser, leveraging the advantages of the three GPU APIs, each adapted to their respective platforms.

The \textit{WebXR Device API}~\cite{webxr} is a working draft developed by the W3C. WebXR standardizes access to \textit{virtual reality} (VR) and \textit{augmented reality} (AR) devices. Interest in VR has been particularly high since easily accessible VR equipment and games reached the market. Multiple use cases can be found for VR and AR, such as GPS navigation, educational content, immersive gaming, among many others that can leverage the technology to provide better content.

Current mobile devices, complemented by a supporting device, are able to provide convincing virtual-reality experiences, prompting game developers to create more content for these systems. Additionally, Meta's announcement of its \textit{Metaverse} in 2021~\cite{metaverse} further supports the need for a VR and AR standard~\cite{wang2022survey}.
There are already a number of promising games that currently make use of the WebXR experimental API, such as \textit{Space Disaster}\footnote{\url{https://www.blend4web.com/apps/space_disaster/space_disaster.html}} or \textit{Moon Rider}.\footnote{\url{https://moonrider.xyz/}} However, it remains to be seen if, or when, these APIs will become popular. The technology and the possibilities are very exciting, but they could still go the way of 3D television, or maybe remorph into something new.

\section{Conclusion}
\label{sec:conclusion}
\hyphenation{WebGPU}
In this paper, we reviewed the past, present, and future trends of browser gaming, highlighting the technologies at hand.
The history of browser-based web games provides a unique perspective on the evolution of the web, and as web browsers evolve and adopt new features, browser game development has grown.
Web games have pushed the boundaries of what browsers can do, leading to the development of new features and, quite importantly, their standardization across platforms.

From the early days of text-based games on the Netscape browser, to modern, 3D MMORPGs, browser games have been developed to account for the limitations and requirements of the era.
Initially, text-based games did not even include graphics due to the lack of support from browsers.
Later, the emergence of DHTML games took advantage of the new dynamic capabilities of browsers to build interactive games.
The introduction of the Java plug-in and applets enabled developers to create complex games, comparable to their native counterparts, while the Flash plug-in empowered an enormous user base to create novel games with advanced animations.
With the advent of HTML5, the browser now natively supports 2D and 3D graphics, allowing for console-grade graphics with WebGL.

Overall, the past and present of browser games have been exciting, and the future looks even more promising as new technologies and standards continue to be developed.
Moreover, the future introduction of WebGPU is set to further reduce, or even eliminate, the gap between browser-based games, native PC games, and game consoles.
Additionally, the \textit{WebXR Device API} standardizes access to virtual and augmented reality devices, which could open up new possibilities for immersive gaming experiences.
With widespread access, browser games have become a popular and ubiquitous form of entertainment and an important part of our culture.
The advancements made in browser game development have not only impacted the gaming industry but have also contributed to the evolution of the web as a whole.

\begin{acks}

We thank Felipe Pepe, Romain Fouquet and Lunar for their invaluable feedback during the writing of this paper.
 
This work has been financially supported by the \textit{Agence Nationale de la Recherche} through the ANR-19-CE39-00201 FP-Locker\footnote{\url{https://anr.fr/Projet-ANR-19-CE39-0002}} and ANR-21-CE39-0019 FACADE\footnote{\url{https://anr.fr/Project-ANR-21-CE39-0019}} projects, and was made possible by \textit{Software Heritage}, \footnote{\url{https://www.softwareheritage.org/}} the great library of source code.

\end{acks}

\clearpage
%\pagebreak
\bibliographystyle{ACM-Reference-Format}
\bibliography{webgames.bib}


\begin{thebibliography}{55}

%%% the Web address.

\ifx \showCODEN    \undefined \def \showCODEN     #1{\unskip}     \fi
\ifx \showDOI      \undefined \def \showDOI       #1{#1}\fi
\ifx \showISBNx    \undefined \def \showISBNx     #1{\unskip}     \fi
\ifx \showISBNxiii \undefined \def \showISBNxiii  #1{\unskip}     \fi
\ifx \showISSN     \undefined \def \showISSN      #1{\unskip}     \fi
\ifx \showLCCN     \undefined \def \showLCCN      #1{\unskip}     \fi
\ifx \shownote     \undefined \def \shownote      #1{#1}          \fi
\ifx \showarticletitle \undefined \def \showarticletitle #1{#1}   \fi
\ifx \showURL      \undefined \def \showURL       {\relax}        \fi
% invisible to TeX
\providecommand\bibfield[2]{#2}
\providecommand\bibinfo[2]{#2}
\providecommand\natexlab[1]{#1}
\providecommand\showeprint[2][]{arXiv:#2}

\bibitem[net(1995)]%
        {netscape2.0b}
 \bibinfo{year}{1995}\natexlab{}.
\newblock \bibinfo{title}{{Netscape Navigator 2.0b1}}.
\newblock
  \bibinfo{howpublished}{\url{https://web.archive.org/web/20020203083536/http://www25.netscape.com:80/eng/mozilla/2.0/relnotes/windows-2.0b1.html}}.
\newblock


\bibitem[spa(1996)]%
        {spaceinvadersAscii}
 \bibinfo{year}{1996}\natexlab{}.
\newblock \bibinfo{title}{{ASCII version of Space Invaders - Reproduced by N.
  Landsteiner}}.
\newblock
  \bibinfo{howpublished}{\url{https://www.masswerk.at/termlib/sample_invaders.html}}.
\newblock


\bibitem[hot(1996)]%
        {hotjavaAlpha}
 \bibinfo{year}{1996}\natexlab{}.
\newblock \bibinfo{title}{{HotJava 1.0 alpha2 (Web Archive)}}.
\newblock
  \bibinfo{howpublished}{\url{https://web.archive.org/web/19961225173659/http://sunsite.unc.edu:80/pub/sun-info/hotjava/}}.
\newblock


\bibitem[int(1996)]%
        {internetExplorer3.0}
 \bibinfo{year}{1996}\natexlab{}.
\newblock \bibinfo{title}{Internet Explorer 3.0 news release}.
\newblock
  \bibinfo{howpublished}{\url{https://news.microsoft.com/1996/08/13/microsoft-launches-microsoft-internet-explorer-3-0-with-exclusive-free-content-offers-from-top-web-sites/}}.
\newblock
\newblock
\shownote{Accessed: 2022-10-30}.


\bibitem[net(1996)]%
        {netscape2.0}
 \bibinfo{year}{1996}\natexlab{}.
\newblock \bibinfo{title}{{Netscape Navigator 2.0 (Web Archive)}}.
\newblock
  \bibinfo{howpublished}{\url{https://web.archive.org/web/19970709120829/http://home.netscape.com/comprod/products/navigator/version_2.0/index.html}}.
\newblock


\bibitem[acq(1997)]%
        {acquisitionFlashMacromedia}
 \bibinfo{year}{1997}\natexlab{}.
\newblock \bibinfo{title}{Macromedia Rides the FutureWave}.
\newblock
  \bibinfo{howpublished}{\url{https://www.wired.com/1997/01/macromedia-rides-the-futurewave/}}.
\newblock
\newblock
\shownote{Accessed: 2022-11-06}.


\bibitem[can(2004)]%
        {canvasWebkit2004}
 \bibinfo{year}{2004}\natexlab{}.
\newblock \bibinfo{title}{Extending HTML}.
\newblock \bibinfo{howpublished}{\url{https://ln.hixie.ch/?start=1089635050}}.
\newblock
\newblock
\shownote{Accessed: 2022-11-07}.


\bibitem[ado(2005)]%
        {adobeMacromedia}
 \bibinfo{year}{2005}\natexlab{}.
\newblock \bibinfo{title}{Adobe Buys Macromedia for \$3.4 Billion}.
\newblock
  \bibinfo{howpublished}{\url{https://www.nytimes.com/2005/04/19/technology/adobe-buys-macromedia-for-34-billion.html}}.
\newblock
\newblock
\shownote{Accessed: 2022-11-06}.


\bibitem[web(2005)]%
        {webApps2005}
 \bibinfo{year}{2005}\natexlab{}.
\newblock \bibinfo{title}{Web Applications 1.0 - Early working draft}.
\newblock
  \bibinfo{howpublished}{\url{https://whatwg.org/specs/web-apps/2005-09-01/}}.
\newblock
\newblock
\shownote{Accessed: 2022-11-06}.


\bibitem[sup(2009)]%
        {supermeatboy}
 \bibinfo{year}{2009}\natexlab{}.
\newblock \bibinfo{title}{SuperMeatBoy Development Blog}.
\newblock
  \bibinfo{howpublished}{\url{http://supermeatboy.blogspot.com/2009/04/i-sleep-to-sounds-of-squrils-making.html}}.
\newblock
\newblock
\shownote{Accessed: 2022-11-07}.


\bibitem[tho(2010)]%
        {thoughtsflash}
 \bibinfo{year}{2010}\natexlab{}.
\newblock \bibinfo{title}{Thoughts on Flash (Web Archive)}.
\newblock
  \bibinfo{howpublished}{\url{https://web.archive.org/web/20100630153444/https://www.apple.com/hotnews/thoughts-on-flash/}}.
\newblock
\newblock
\shownote{Accessed: 2022-11-03}.


\bibitem[har(2012)]%
        {hardwareAcc}
 \bibinfo{year}{2012}\natexlab{}.
\newblock \bibinfo{title}{Taking advantage of GPU acceleration in the 2D
  canvas}.
\newblock
  \bibinfo{howpublished}{\url{https://developer.chrome.com/blog/taking-advantage-of-gpu-acceleration-in-the-2d-canvas/}}.
\newblock


\bibitem[end(2015)]%
        {endNpapiUnity}
 \bibinfo{year}{2015}\natexlab{}.
\newblock \bibinfo{title}{Mozilla - NPAPI Plugins in Firefox}.
\newblock
  \bibinfo{howpublished}{\url{https://blog.mozilla.org/futurereleases/2015/10/08/npapi-plugins-in-firefox/}}.
\newblock


\bibitem[end(2017)]%
        {endoflife}
 \bibinfo{year}{2017}\natexlab{}.
\newblock \bibinfo{title}{Flash \& the Future of Interactive Content}.
\newblock
  \bibinfo{howpublished}{\url{https://blog.adobe.com/en/publish/2017/07/25/adobe-flash-update
  }}.
\newblock
\newblock
\shownote{Accessed: 2022-11-07}.


\bibitem[unr(2019)]%
        {unrealHtml5}
 \bibinfo{year}{2019}\natexlab{}.
\newblock \bibinfo{title}{Unreal Engine - HTML5 Game Development}.
\newblock
  \bibinfo{howpublished}{\url{https://docs.unrealengine.com/4.27/en-US/SharingAndReleasing/HTML5/}}.
\newblock
\newblock
\shownote{Accessed: 2022-11-06}.


\bibitem[fla(2022)]%
        {flashpoint}
 \bibinfo{year}{2022}\natexlab{}.
\newblock \bibinfo{title}{Bluemaxima's Flashpoint}.
\newblock \bibinfo{howpublished}{\url{https://bluemaxima.org/flashpoint/}}.
\newblock
\newblock
\shownote{Accessed: 2022-11-05}.


\bibitem[dir(2022)]%
        {directx}
 \bibinfo{year}{2022}\natexlab{}.
\newblock \bibinfo{title}{Direct3D 12 Api}.
\newblock
  \bibinfo{howpublished}{\url{https://docs.microsoft.com/en-us/windows/win32/direct3d12/direct3d-12-graphics}}.
\newblock


\bibitem[khr(2022)]%
        {khronos}
 \bibinfo{year}{2022}\natexlab{}.
\newblock \bibinfo{title}{Khronos Group}.
\newblock \bibinfo{howpublished}{\url{https://www.khronos.org/}}.
\newblock


\bibitem[ear(2022a)]%
        {earningsmicrosoft}
 \bibinfo{year}{2022}\natexlab{a}.
\newblock \bibinfo{title}{Microsoft First Quarter Earnings Conference Call}.
\newblock
  \bibinfo{howpublished}{\url{https://www.microsoft.com/en-us/Investor/events/FY-2023/earnings-fy-2023-q1.aspx}}.
\newblock
\newblock
\shownote{Accessed: 2022-11-06}.


\bibitem[mos(2022)]%
        {mosaicBrowser}
 \bibinfo{year}{2022}\natexlab{}.
\newblock \bibinfo{title}{Mosaic browser}.
\newblock
  \bibinfo{howpublished}{\url{https://www.ncsa.illinois.edu/research/project-highlights/ncsa-mosaic/}}.
\newblock
\newblock
\shownote{Accessed: 2022-10-31}.


\bibitem[new(2022)]%
        {newsground}
 \bibinfo{year}{2022}\natexlab{}.
\newblock \bibinfo{title}{Newgrounds}.
\newblock \bibinfo{howpublished}{\url{https://www.newgrounds.com/}}.
\newblock
\newblock
\shownote{Accessed: 2022-11-05}.


\bibitem[int(2022)]%
        {internetaccess}
 \bibinfo{year}{2022}\natexlab{}.
\newblock \bibinfo{title}{OECD - Broadband statistics}.
\newblock
  \bibinfo{howpublished}{\url{https://www.oecd.org/sti/broadband/broadband-statistics/}}.
\newblock


\bibitem[pha(2022)]%
        {phaser}
 \bibinfo{year}{2022}\natexlab{}.
\newblock \bibinfo{title}{Phaser Game Engine}.
\newblock \bibinfo{howpublished}{\url{https://phaser.io/}}.
\newblock


\bibitem[son(2022)]%
        {sony}
 \bibinfo{year}{2022}\natexlab{}.
\newblock \bibinfo{title}{Sony Corporation}.
\newblock \bibinfo{howpublished}{\url{https://sony.com}}.
\newblock


\bibitem[ear(2022b)]%
        {earningsnvidia}
 \bibinfo{year}{2022}\natexlab{b}.
\newblock \bibinfo{title}{Transcript - Nvidia's earnings call}.
\newblock
  \bibinfo{howpublished}{\url{https://www.fool.com/earnings/call-transcripts/2022/08/24/nvidia-nvda-q2-2023-earnings-call-transcript/}}.
\newblock
\newblock
\shownote{Accessed: 2022-11-06}.


\bibitem[uni(2022)]%
        {unityGE}
 \bibinfo{year}{2022}\natexlab{}.
\newblock \bibinfo{title}{Unity}.
\newblock \bibinfo{howpublished}{\url{https://unity.com}}.
\newblock


\bibitem[unr(2022)]%
        {unrealGE}
 \bibinfo{year}{2022}\natexlab{}.
\newblock \bibinfo{title}{Unreal Engine}.
\newblock \bibinfo{howpublished}{\url{https://www.unrealengine.com}}.
\newblock


\bibitem[vul(2022)]%
        {vulkan}
 \bibinfo{year}{2022}\natexlab{}.
\newblock \bibinfo{title}{Vulkan API}.
\newblock \bibinfo{howpublished}{\url{https://www.vulkan.org/}}.
\newblock


\bibitem[web(2022)]%
        {webgpu}
 \bibinfo{year}{2022}\natexlab{}.
\newblock \bibinfo{title}{W3C - WebGPU working draft}.
\newblock \bibinfo{howpublished}{\url{https://www.w3.org/TR/webgpu/}}.
\newblock


\bibitem[far(2023)]%
        {farmville}
 \bibinfo{year}{2023}\natexlab{}.
\newblock \bibinfo{title}{Farmville}.
\newblock \bibinfo{howpublished}{\url{https://www.farmville3.com/}}.
\newblock


\bibitem[neo(2023)]%
        {neopets}
 \bibinfo{year}{2023}\natexlab{}.
\newblock \bibinfo{title}{Neopets}.
\newblock \bibinfo{howpublished}{\url{https://www.neopets.com/}}.
\newblock


\bibitem[sta(2023)]%
        {stardoll}
 \bibinfo{year}{2023}\natexlab{}.
\newblock \bibinfo{title}{Stardoll}.
\newblock \bibinfo{howpublished}{\url{https://www.stardoll.com/}}.
\newblock


\bibitem[{Apple}(2022)]%
        {metal}
\bibfield{author}{\bibinfo{person}{{Apple}}.} \bibinfo{year}{2022}\natexlab{}.
\newblock \bibinfo{title}{Metal API}.
\newblock \bibinfo{howpublished}{\url{https://developer.apple.com/metal/}}.
\newblock


\bibitem[Buhov et~al\mbox{.}(2018)]%
        {buhov2018flash}
\bibfield{author}{\bibinfo{person}{Damjan Buhov}, \bibinfo{person}{Julian
  Rauchberger}, {and} \bibinfo{person}{Sebastian Schrittwieser}.}
  \bibinfo{year}{2018}\natexlab{}.
\newblock \showarticletitle{FLASH: Is the 20th Century Hero Really Gone?
  Large-Scale Evaluation on Flash Usage \& Its Security and Privacy
  Implications.}
\newblock \bibinfo{journal}{\emph{J. Wirel. Mob. Networks Ubiquitous Comput.
  Dependable Appl.}} \bibinfo{volume}{9}, \bibinfo{number}{4}
  (\bibinfo{year}{2018}), \bibinfo{pages}{26--40}.
\newblock


\bibitem[Cai et~al\mbox{.}(2016)]%
        {cai2016survey}
\bibfield{author}{\bibinfo{person}{Wei Cai}, \bibinfo{person}{Ryan Shea},
  \bibinfo{person}{Chun-Ying Huang}, \bibinfo{person}{Kuan-Ta Chen},
  \bibinfo{person}{Jiangchuan Liu}, \bibinfo{person}{Victor~CM Leung}, {and}
  \bibinfo{person}{Cheng-Hsin Hsu}.} \bibinfo{year}{2016}\natexlab{}.
\newblock \showarticletitle{A survey on cloud gaming: Future of computer
  games}.
\newblock \bibinfo{journal}{\emph{IEEE Access}}  \bibinfo{volume}{4}
  (\bibinfo{year}{2016}), \bibinfo{pages}{7605--7620}.
\newblock


\bibitem[Chen et~al\mbox{.}(2011)]%
        {chen2011measuring}
\bibfield{author}{\bibinfo{person}{Kuan-Ta Chen}, \bibinfo{person}{Yu-Chun
  Chang}, \bibinfo{person}{Po-Han Tseng}, \bibinfo{person}{Chun-Ying Huang},
  {and} \bibinfo{person}{Chin-Laung Lei}.} \bibinfo{year}{2011}\natexlab{}.
\newblock \showarticletitle{Measuring the latency of cloud gaming systems}. In
  \bibinfo{booktitle}{\emph{Proceedings of the 19th ACM international
  conference on Multimedia}}. \bibinfo{pages}{1269--1272}.
\newblock


\bibitem[Chiaravutthi(2006)]%
        {chiaravutthi2006firms}
\bibfield{author}{\bibinfo{person}{Yingyot Chiaravutthi}.}
  \bibinfo{year}{2006}\natexlab{}.
\newblock \showarticletitle{Firms' strategies and network externalities:
  Empirical evidence from the browser war}.
\newblock \bibinfo{journal}{\emph{The Journal of High Technology Management
  Research}} \bibinfo{volume}{17}, \bibinfo{number}{1} (\bibinfo{year}{2006}),
  \bibinfo{pages}{27--42}.
\newblock


\bibitem[Curtis(2014)]%
        {curtis2014online}
\bibfield{author}{\bibinfo{person}{Vickie Curtis}.}
  \bibinfo{year}{2014}\natexlab{}.
\newblock \showarticletitle{Online citizen science games: opportunities for the
  biological sciences}.
\newblock \bibinfo{journal}{\emph{Applied \& translational genomics}}
  \bibinfo{volume}{3}, \bibinfo{number}{4} (\bibinfo{year}{2014}),
  \bibinfo{pages}{90--94}.
\newblock


\bibitem[Fiadotau(2020)]%
        {fiadotau2020growing}
\bibfield{author}{\bibinfo{person}{Mikhail Fiadotau}.}
  \bibinfo{year}{2020}\natexlab{}.
\newblock \showarticletitle{Growing old on Newgrounds: The hopes and quandaries
  of Flash game preservation}.
\newblock \bibinfo{journal}{\emph{First Monday}} (\bibinfo{year}{2020}).
\newblock


\bibitem[Guldman(2002)]%
        {guldman2002building}
\bibfield{author}{\bibinfo{person}{Andrew Guldman}.}
  \bibinfo{year}{2002}\natexlab{}.
\newblock \showarticletitle{Building Rich Internet Applications with Macromedia
  Flash MX and ColdFusion MX}.
\newblock \bibinfo{journal}{\emph{Macromedia white paper}}
  (\bibinfo{year}{2002}).
\newblock


\bibitem[Hawkes(2011)]%
        {hawkes2011foundation}
\bibfield{author}{\bibinfo{person}{Rob Hawkes}.}
  \bibinfo{year}{2011}\natexlab{}.
\newblock \bibinfo{booktitle}{\emph{Foundation HTML5 Canvas: For Games and
  Entertainment}}.
\newblock \bibinfo{publisher}{Apress}.
\newblock


\bibitem[{HP}(2021)]%
        {bestSellingHP}
\bibfield{author}{\bibinfo{person}{{HP}}.} \bibinfo{year}{2021}\natexlab{}.
\newblock \bibinfo{title}{Top 50 best-selling video games of all time}.
\newblock
  \bibinfo{howpublished}{\url{https://www.hp.com/us-en/shop/tech-takes/top-50-best-selling-video-games-all-time}}.
\newblock
\newblock
\shownote{Accessed: 2022-11-05}.


\bibitem[Keith(2005)]%
        {keith2005brief}
\bibfield{author}{\bibinfo{person}{Jeremy Keith}.}
  \bibinfo{year}{2005}\natexlab{}.
\newblock \showarticletitle{A Brief History of JavaScript}.
\newblock \bibinfo{journal}{\emph{DOM Scripting: Web Design with JavaScript and
  the Document Object Model}} (\bibinfo{year}{2005}), \bibinfo{pages}{3--10}.
\newblock


\bibitem[Lujan et~al\mbox{.}(2019)]%
        {lujan2019evaluating}
\bibfield{author}{\bibinfo{person}{Michael Lujan}, \bibinfo{person}{Michael
  Baum}, \bibinfo{person}{Dayuan Chen}, {and} \bibinfo{person}{Ziliang Zong}.}
  \bibinfo{year}{2019}\natexlab{}.
\newblock \showarticletitle{Evaluating the Performance and Energy Efficiency of
  OpenGL and Vulkan on a Graphics Rendering Server}. In
  \bibinfo{booktitle}{\emph{2019 International Conference on Computing,
  Networking and Communications (ICNC)}}. IEEE, \bibinfo{pages}{777--781}.
\newblock


\bibitem[{Meta}(2021)]%
        {metaverse}
\bibfield{author}{\bibinfo{person}{{Meta}}.} \bibinfo{year}{2021}\natexlab{}.
\newblock \bibinfo{title}{Our vision for the metaverse}.
\newblock
  \bibinfo{howpublished}{\url{https://tech.fb.com/ar-vr/2021/10/connect-2021-our-vision-for-the-metaverse/}}.
\newblock
\newblock
\shownote{Accessed: 2022-11-06}.


\bibitem[{Mozilla}(2022)]%
        {webxr}
\bibfield{author}{\bibinfo{person}{{Mozilla}}.}
  \bibinfo{year}{2022}\natexlab{}.
\newblock \bibinfo{title}{WebXR Device API}.
\newblock
  \bibinfo{howpublished}{\url{https://developer.mozilla.org/en-US/docs/Web/API/WebXR_Device_API}}.
\newblock
\newblock
\shownote{Accessed: 2022-11-06}.


\bibitem[{Oracle}(2017)]%
        {jdk9Deprecated}
\bibfield{author}{\bibinfo{person}{{Oracle}}.} \bibinfo{year}{2017}\natexlab{}.
\newblock \bibinfo{title}{JDK 9 Release Notes}.
\newblock
  \bibinfo{howpublished}{\url{https://www.oracle.com/java/technologies/javase/9-deprecated-features.html}}.
\newblock
\newblock
\shownote{Accessed: 2022-11-04}.


\bibitem[Patrick(2019)]%
        {patrick2019brief}
\bibfield{author}{\bibinfo{person}{Michael~D Patrick}.}
  \bibinfo{year}{2019}\natexlab{}.
\newblock \showarticletitle{A Brief History of Digital Communications}.
\newblock In \bibinfo{booktitle}{\emph{Social Media for Medical
  Professionals}}. \bibinfo{publisher}{Springer}, \bibinfo{pages}{23--47}.
\newblock


\bibitem[Salter and Murray(2014)]%
        {salter2014flash}
\bibfield{author}{\bibinfo{person}{Anastasia Salter} {and}
  \bibinfo{person}{John Murray}.} \bibinfo{year}{2014}\natexlab{}.
\newblock \bibinfo{booktitle}{\emph{Flash: Building the interactive web}}.
\newblock \bibinfo{publisher}{MIT Press}.
\newblock


\bibitem[Statista(2020)]%
        {gameindustry}
\bibfield{author}{\bibinfo{person}{Statista}.} \bibinfo{year}{2020}\natexlab{}.
\newblock \bibinfo{title}{Gaming: The Most Lucrative Entertainment Industry By
  Far}.
\newblock
  \bibinfo{howpublished}{\url{https://www.thc-pod.com/episode/the-gaming-industry-is-now-bigger-than-movies-and-music-combined}}.
\newblock
\newblock
\shownote{Accessed: 2022-11-08}.


\bibitem[Stobart and Vassileiou(2004)]%
        {stobart2004introduction}
\bibfield{author}{\bibinfo{person}{Simon Stobart} {and} \bibinfo{person}{Mike
  Vassileiou}.} \bibinfo{year}{2004}\natexlab{}.
\newblock \showarticletitle{Introduction to PHP}.
\newblock In \bibinfo{booktitle}{\emph{PHP and MySQL Manual}}.
  \bibinfo{publisher}{Springer}, \bibinfo{pages}{7--11}.
\newblock


\bibitem[W3C(1996)]%
        {w3chtml4}
\bibfield{author}{\bibinfo{person}{W3C}.} \bibinfo{year}{1996}\natexlab{}.
\newblock \bibinfo{title}{HTML 4.0 press release}.
\newblock
  \bibinfo{howpublished}{\url{https://www.w3.org/Press/HTML4-REC-fact.html}}.
\newblock
\newblock
\shownote{Accessed: 2022-11-06}.


\bibitem[Wang et~al\mbox{.}(2022)]%
        {wang2022survey}
\bibfield{author}{\bibinfo{person}{Yuntao Wang}, \bibinfo{person}{Zhou Su},
  \bibinfo{person}{Ning Zhang}, \bibinfo{person}{Rui Xing},
  \bibinfo{person}{Dongxiao Liu}, \bibinfo{person}{Tom~H Luan}, {and}
  \bibinfo{person}{Xuemin Shen}.} \bibinfo{year}{2022}\natexlab{}.
\newblock \showarticletitle{A survey on metaverse: Fundamentals, security, and
  privacy}.
\newblock \bibinfo{journal}{\emph{IEEE Communications Surveys \& Tutorials}}
  (\bibinfo{year}{2022}).
\newblock


\bibitem[Whitson and Dormann(2011)]%
        {whitson2011social}
\bibfield{author}{\bibinfo{person}{Jennifer~R Whitson} {and}
  \bibinfo{person}{Claire Dormann}.} \bibinfo{year}{2011}\natexlab{}.
\newblock \showarticletitle{Social gaming for change: Facebook unleashed}.
\newblock \bibinfo{journal}{\emph{First Monday}} (\bibinfo{year}{2011}).
\newblock


\bibitem[Wirfs-Brock and Eich(2020)]%
        {brock2020javascript}
\bibfield{author}{\bibinfo{person}{Allen Wirfs-Brock} {and}
  \bibinfo{person}{Brendan Eich}.} \bibinfo{year}{2020}\natexlab{}.
\newblock \showarticletitle{JavaScript: The First 20 Years}.
\newblock \bibinfo{journal}{\emph{Proc. ACM Program. Lang.}}
  \bibinfo{volume}{4}, \bibinfo{number}{HOPL}, Article \bibinfo{articleno}{77}
  (\bibinfo{date}{jun} \bibinfo{year}{2020}), \bibinfo{numpages}{189}~pages.
\newblock
\urldef\tempurl%
\url{https://doi.org/10.1145/3386327}
\showDOI{\tempurl}


\end{thebibliography}

\clearpage
\onecolumn

% %\clearpage
% %\pagebreak
% \appendix
% \section{Server power measurements}
% \label{apdx:powermeasures}
% Figure~\ref{fig:power_measures} depicts the average power consumption of various backend Web stacks based on the number of simultaneous clients. Power consumption increases with the number of clients. However, for most Web servers, the average power consumption tends to plateau after 32 simultaneous clients.

% \begin{figure*}[h]
%   \centering
%   \includegraphics[width=1.0\textwidth]{figures/avg_power_backends.pdf}
%   \caption{Average power consumed per number of clients for each backend stack}
%   \label{fig:power_measures}
% \end{figure*}

\end{document}